\documentclass[runningheads]{llncs}
\usepackage{array}
\usepackage[table,xcdraw]{xcolor}
\usepackage[T1]{fontenc}
\usepackage{amssymb}
\usepackage{wrapfig}
\usepackage{subcaption}
\usepackage{listings}
\usepackage{pgfplots}
\pgfplotsset{width=12cm,compat=1.16}
\usepackage{graphicx}
\usepackage{hyperref}
\usepackage{color}
\usepackage{subcaption}
\usepackage{soul}

\usepackage{makecell}
\usepackage{wrapfig}
\usepackage[inline]{enumitem}
\usepackage{comment}
\usepackage{tikz}
\usepackage{todonotes}

\usetikzlibrary{arrows,automata,positioning}
\begin{document}
\title{Reusable Formal Verification\\ of DAG-based Consensus Protocols}
%

%
\author{Nathalie Bertrand\inst{1}\orcidID{0000-0002-9957-5394} \and
Pranav Ghorpade\inst{2}\orcidID{0009-0001-0421-4490} \and
Sasha Rubin\inst{2}\orcidID{0000-0002-3948-129X} \and
Bernhard Scholz\inst{3}\orcidID{0000-0002-7672-7359} \and
Pavle Suboti\'c\inst{3}\orcidID{0000-0002-6536-3932}
}
\authorrunning{N. Bertrand, P. Ghorpade, S. Rubin, B. Scholz and P. Subotić}
%
\institute{Univ Rennes, Inria, CNRS, IRISA \and The University of Sydney \and Sonic Research }
\maketitle              

\lstset{language=Python, basicstyle=\scriptsize\ttfamily,breaklines=true,
  backgroundcolor=\color{gray!10}, showstringspaces=false
}

\begin{abstract}
  
Blockchains use consensus protocols to reach agreement, e.g., on the ordering of transactions. DAG-based consensus protocols are increasingly adopted by blockchain companies to reduce energy consumption and enhance security. These protocols collaboratively construct a partial order of blocks (DAG construction) and produce a linear sequence of blocks (DAG ordering). Given the strategic significance of blockchains, formal proofs of the correctness of key components such as consensus protocols are essential. This paper presents safety-verified specifications for five DAG-based consensus protocols. Four of these protocols ---DAG-Rider, Cordial Miners, Hashgraph, and Eventual Synchronous BullShark --- are well-established in the literature. The fifth protocol is a minor variation of Aleph, another well-established protocol. Our framework enables proof reuse, reducing proof efforts by almost half. It achieves this by providing various independent, formally verified, specifications of DAG construction and ordering variations, which can be combined to express all five protocols. We employ TLA+ for specifying the protocols and writing their proofs, and the TLAPS proof system to automatically check the proofs. Each TLA+ specification is relatively compact, and TLAPS efficiently verifies hundreds to thousands of obligations within minutes. The significance of our work is two-fold: first, it supports the adoption of DAG-based systems by providing robust safety assurances; second, it illustrates that DAG-based consensus protocols are amenable to practical, reusable, and compositional formal methods.

\end{abstract}

\section{Introduction}
\label{sec:intro}
At the core of cryptocurrency lies blockchain technology, which relies on consensus protocols to coordinate network processes to achieve agreement on the state of the blockchain.
Early blockchain consensus protocols, such as those used in Bitcoin and Tendermint, depend on varying degrees of synchrony within their operational environments to ensure safety (preventing conflicting transactions) and liveness (guaranteeing eventual transaction confirmation)~\cite{bitcoin,Buterin2013,tendermint,algorand}.
However, recent advances have introduced asynchronous probabilistic consensus protocols based on Directed Acyclic Graphs (DAGs) ~\cite{KKNS-podc21,KeidarNPS23,BL-coins20,GagolLSS19,lachesis,Giridharan2022BullsharkDB}.
These protocols not only demonstrate high performance and guarantee Byzantine Fault Tolerance (BFT) but also utilize processes fairly and exhibit low communication complexity. 
Given these advantages, there has been a growing interest in DAG-based protocols in both industrial and academic circles \cite{sokDAGcons}. 
Several leading blockchains have adopted DAG-based protocols as their primary consensus mechanisms~\cite{aptos,lachesis,hedra}. 
Given the trillions of dollars locked in various blockchains~\cite{tlv}, manipulating the consensus protocol of the blockchain is a natural attack vector.
Double-spending attacks~\cite{doublespending} exploit unsafe protocols by identifying inputs and conditions that lead to inconsistent blockchain states, allowing a currency unit to be spent multiple times.
Although consensus protocols are generally designed to be safe and thus mitigate such attacks, ad hoc software design practices complicate the assurance of safety across all potential inputs. For example, even well-established, state-of-the-art blockchains are unsafe~\cite{attackseth,shoup}.
This underscores the need for solutions that ensure safety in the design of consensus protocols.

A common approach involves testing. However, a testing regime for a software implementation might not reveal deficiencies in the protocol design itself, and malicious behaviors of byzantine processes may not be fully understood and covered by test cases. Another approach is model checking, but model checking frequently encounters the state-space explosion problem, making it infeasible to exhaustively verify models as they scale. Together, both testing and traditional model checking cannot provide guarantees of correctness no matter the number of participating processes and no matter their configurations and behavior. Given the vast number of possible interleavings in an asynchronous environment, omitting rigorous verification is risky.

A more robust method for ensuring safety is \emph{formal proofs}: constructing a mathematical model that describes the system behavior and providing a mathematical proof that the model is correct for all possible inputs and configurations.
While such proof efforts exist for a small number of consensus protocols~\cite{moonshotverif,ChandLS16,SchultzDT22,BraithwaiteB0MS20,redbellyverif,DBLP:conf/vmcai/AminofRSWZ18,Thomsen2020FormalizingNP,10.1145/3167086,10.1145/2815400.2815428}, the vast majority of consensus protocols operate with an informal assumption of safety.
This is often due to perception that formal proofs are tedious, time-consuming, and complex, largely because of a lack of compositional \emph{building blocks} that facilitate the proof of correctness for new protocols.
While reuse is often achievable at the specification level, the corresponding safety proofs generally lack compositionality, making verification challenging.

In this paper, we present an industrial case study where proven models of consensus protocols are used to drive development. We emphasize making formal proofs more practical, reusable, and compositional for distributed systems that leverage DAG-based consensus protocols. 
This is achieved by creating compositional specifications and proofs that encapsulate commonalities between protocols, thereby allowing the reuse of proofs. Our overall approach relies on the principle of \emph{equality by abstraction}~\cite{abstraction1}: by selecting appropriate abstractions for DAG-based protocols, we can decompose protocol properties and proofs into modular components.
Using this approach we provide safety-verified specifications for five DAG-based protocols DAG-Rider~\cite{KKNS-podc21}, Cordial Miners~\cite{KeidarNPS23}, Hashgraph~\cite{BL-coins20}, Eventually Synchronous BullShark~\cite{Giridharan2022BullsharkDB}, and a variant of Aleph. 
Our experience shows that while these (and many other) DAG-based protocols share fundamental principles, they differ in the variations introduced to improve performance. Nonetheless, these variations can significantly impact their correctness. Thus, our work demonstrates significant potential for reuse in handling other DAG based consensus protocols~\cite{sokDAGcons}.
We implemented an open-source formal specification of above mentioned protocols (available at~\cite{anonrepo}). 
Our specification and proofs are in the Temporal Logic of Actions (TLA)~\cite{Lamport-book2002}. For this we use TLA+, tool to write specification and proofs in TLA.
We check proofs using the TLA+ proof system (TLAPS)~\cite{CDLMRV-fm12}, a proof system for mechanically checking proofs written in TLA. 
We use TLA+ as it is a popular specification language with numerous industrial case studies~\cite{tlaamazon,tlams} and well supported tooling~\cite{Konnov}. 
\section{DAG-based Consensus Protocols in Blockchains}
\label{sec:dagprotocols}\label{sec:background}

In this section, we synthesize the fundamental principles shared by DAG-based consensus protocols, as well as common variations introduced to enhance performance.

DAG-based consensus protocols~\cite{KKNS-podc21,KeidarNPS23,BL-coins20,GagolLSS19,lachesis,Giridharan2022BullsharkDB} operate in a distributed environment consisting of $n$ processes that communicate through reliable peer-to-peer message channels, meaning that messages are neither lost nor duplicated. 
Each process is assumed to have an id and all messages are signed by their sender, ensuring that they cannot be forged.
Up to $f < \frac{n}{3}$ of the processes are Byzantine-faulty, which means that they may deviate from the protocol.
Such processes can send any message to others or refuse to send certain messages. 
Correct processes, in contrast, adhere to the algorithm’s specification and continue execution indefinitely.
The system is typically asynchronous, meaning that processes operate at independent speeds and messages may experience arbitrary delays. However, there are exceptions, e.g.,  Eventual Synchronous Bullshark (ES Bullshark) model assumes certain bounds on message delays \cite{Giridharan2022BullsharkDB}.

DAG-based consensus protocols solve the Byzantine Atomic Broadcast (BAB) problem~\cite{6076782,8416469,CRISTIAN1995158}. 
BAB provides a mechanism to propose sets of transactions (blocks) and totally order them in blockchain systems with Byzantine faulty processes.
Informally, Byzantine Atomic Broadcast guarantees the following:
\begin{enumerate}
    \item \textbf{Agreement:} If a correct process delivers a message, then all other correct processes eventually deliver that message with probability 1.
    \item \textbf{Validity:} All messages broadcast by correct processes are eventually delivered with probability $1$. \footnote{A probabilistic guarantee is needed due to asynchronous nature of the communication and Fischer, Lynch, and Paterson (FLP) impossibility result~\cite{FLPjacm85}.}
    \item \textbf{Integrity:} A correct process delivers a message at most once, and only if it was previously broadcast. 
    \item \textbf{Total Order:} All correct processes agree on the delivery order of the set of messages they deliver.
\end{enumerate}

Although we have not found an agreed-upon definition of ``DAG-based consensus protocols'', we propose an informal definition that matches many published protocols~\cite{KKNS-podc21,KeidarNPS23,BL-coins20,GagolLSS19,lachesis,Giridharan2022BullsharkDB}, further detailed in Figure \ref{fig:surveyDAG}:\\

\noindent\fbox{%
    \parbox{\textwidth}{%
        DAG-based consensus protocols solve the BAB problem in two phases:
        (1) In the \textbf{DAG construction phase} processes communicate their blocks and construct a Directed Acyclic Graph (DAG) of the exchanged blocks. (2) In the \textbf{ordering phase}, processes use their locally constructed DAGs to determine a total order of the blocks without requiring additional communication.
    }%
    \label{twophaseDAG}
}\\

In the rest of this section, we broadly outline our view of these two phases. We do this in enough detail that a reader who wants to model their own DAG-based consensus protocol can appreciate how to separate their protocol into the two phases, and how to apply our abstractions in Section~\ref{sec:abs}.
Our discussion, however, is restricted to how DAG-based consensus protocols ensure Integrity and Total Order --- the safety properties we aim to verify.
We do not focus on liveness properties such as Validity or the eventual delivery aspect of Agreement, which are outside the scope of this work.

\subsection{DAG Construction Phase} \label{background:DAGcon}

In the DAG construction phase, processes create and communicate blocks in the form of vertices. 
A vertex is defined recursively and consists of three components: the creator’s ID, the block the creator proposes, and a possibly empty set of references to other vertices.
Each process $p$ builds a Directed Acyclic Graph (DAG) $G(p)$, called its \emph{local DAG}, which includes vertices created by process $p$, as well as some of those received from other processes. References act as outgoing edges: if $v,v'$ are vertices in $G(p)$ and $v'$ is referenced by $v$, then $G(p)$ has a directed edge from $v$ to $v'$.
Vertices are added sequentially to $G(p)$. The key property of the construction of $G(p)$ is that a newly created or received vertex $v$ is only added to $G(p)$ once all the vertices referenced by $v$ have been added to $G(p)$ (in the meantime, $v$ is stored in a buffer).
Thus, $G(p)$ is indeed acyclic: if a vertex $v$ is added to $G(p)$, and if there is a reference path from $v$ to some vertex $v'$, then $v'$ was previously added to $G(p)$.
Note that the local DAGs of two correct processes need not be identical or even subgraphs of each other. However they satisfy following property: 
\begin{property}[Consistent Causal History]\label{property1}
If a vertex $v$ appears in the local DAGs of any two correct processes then the sub-DAG rooted at $v$, also called the \emph{causal history} of $v$, is identical in both.
\end{property}
Variations in the DAG construction phase in different protocols arise from two main factors: the communication primitive used to exchange blocks and the type of DAG being constructed (see Fig. \ref{fig:surveyDAG}).

\paragraph{Communication.}
Some protocols, e.g., DAG-Rider, BullShark, and Aleph, use a \emph{reliable broadcast} primitive to communicate vertices (the specification of Reliable broadcast is the same as that of Byzantine Atomic Broadcast but without Total Order). 
As a result, these protocols prevent equivocation, meaning that no vertex appears more than once in the local DAGs of correct processes.
In contrast, other protocols such as Cordial Miners, Hashgraph, and Lachesis, rely on unreliable communication to 
spread vertices, e.g., plain broadcast~\cite{KeidarNPS23} or gossip~\cite{gossip}. 
While this approach may allow Byzantine processes to introduce conflicting vertices, it offers significantly lower latency compared to reliable broadcast.

\paragraph{DAG Type.}  
In protocols like DAG-Rider, Cordial Miners, BullShark, and Aleph, processes create new vertices in rounds, one in each. Thus each vertex also contains its creation round number.
A process has \emph{completed} round $r$ once its local DAG contains vertices created by at least $n-f$ processes in round $r$. 
To create a vertex $v$ in round $r > 0$, a process must wait to complete round $r-1$ and ensure that $v$ references all the vertices in its local DAG from  round $r-1$ at the time of creation. 
This round-based construction results in a \emph{layered structure}: vertices in round $r = 0$  form the first (bottom) layer, while vertices in round $r > 0$ have edges to at least $n-f$ vertices in round $r-1$ created by distinct processes. 
Figure~\ref{fig:dis} illustrates this structure for DAG-Rider. 

On the other hand, in Hashgraph and Lachesis there is no use of rounds. Process $p$ creates a new vertex $v$ upon receiving some vertex $v'$ from another process. Vertex $v$ references exactly two vertices: the last vertex created by $p$, and $v'$.
As a result, these DAGs lack a layered structure as that of previous protocols.
While ordering protocols are simpler to implement on layered structured DAGs, unstructured DAGs capture a finer partial order of vertices, helping provide additional {fairness} guarantees \cite{BL-coins20}.

As discussed above, there are performance trade-offs in the way local DAGs are constructed. While we have outlined some common configurations, the design space offers numerous possibilities. For instance, one could combine reliable and unreliable communication to balance the trade-off between latency and fault tolerance when exchanging blocks \cite{lightdag}.

\subsection{Ordering Phase} \label{background:Ordering}

Each process maintains a local DAG, representing a partial order of vertices,  that grows during the DAG construction phase. The ordering phase takes this evolving DAG as input (assumed to satisfy the Consistent Causal History Property), and is responsible for totally ordering a subset of its vertices for final delivery. This must be achieved without additional communication while ensuring that all correct processes deliver vertices in the same order (Total Order) and do not deliver duplicate vertices more than once (Integrity). We now describe a procedure that guarantees {Total Order} and {Integrity}, applicable to all protocols discussed in this paper.

Each process partitions the vertices in its local DAG  into sequentially numbered ``frames'', starting from \(0\). As the DAG grows, new frames are introduced. For example, in a structured DAG, a frame corresponds to a fixed number of consecutive rounds (e.g., 4 for DAG Rider), often referred to as a ``wave''.
Every correct process $p$, for each frame $x$, periodically attempts to select a subset of vertices $A_{(p, x)}$ from those currently in frame $x$ of $p's$ local DAG to serve as ``anchors'', so that there is an agreement on anchors across correct process, as formalized below:
\begin{property}[Anchor Agreement]\label{property2}
    For a correct process \( p \) and frame \( x \), let \( A_{(p, x)} \) be the set of anchors selected by process \( p \) for frame \( x \), initially set to \( \textit{Nil} \) to indicate that anchors have not yet been chosen. Then, for every frame \( x \) and any two correct processes \( p, q \), if \( A_{(p, x)} \neq \textit{Nil} \) and \( A_{(q, x)} \neq \textit{Nil} \), then \( A_{(p, x)} = A_{(q, x)} \).
\end{property}
Since anchor selection depends on the available information, i.e., the current local DAG, a frame's anchor need not be selected when the frame was defined; it might be selected in a later period after the local DAG has evolved.
Different protocols employ different mechanisms to achieve agreement on anchors, as we discuss at the end of this subsection.

\begin{figure*}[t]
    \centering
    \begin{subfigure}[t]{0.45\textwidth}
        \centering
        \includegraphics[width=\textwidth]{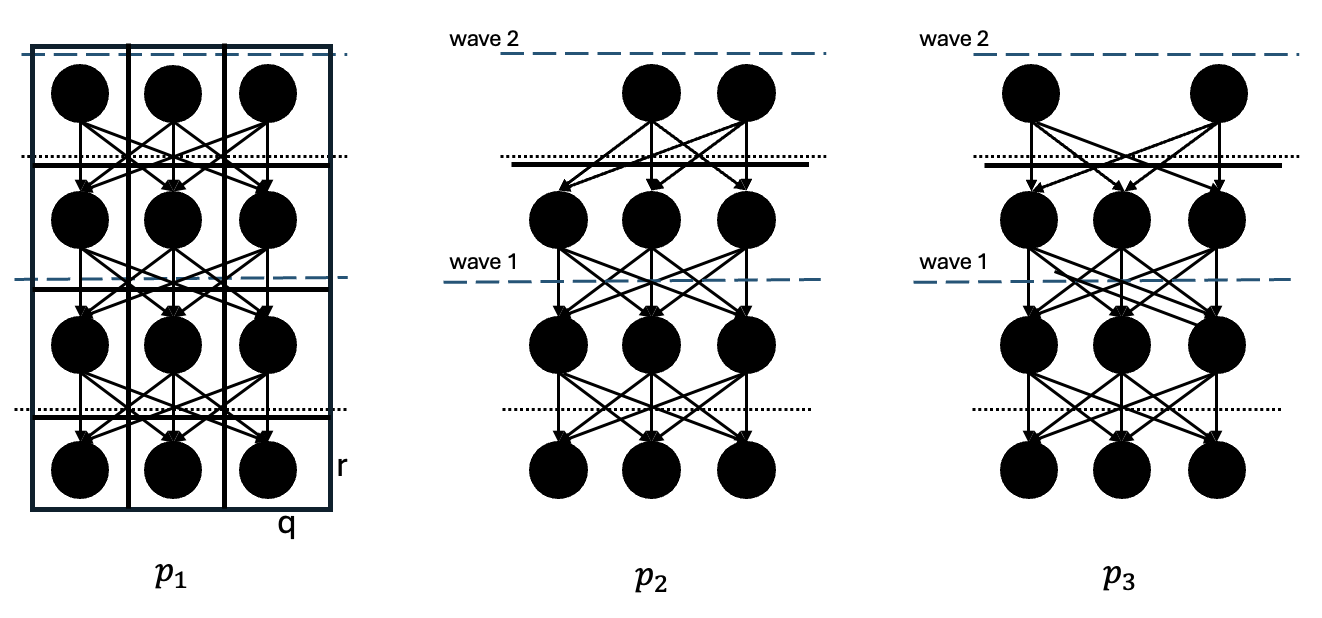}
        \caption{DAG construction phase: building local DAGs for each
          processes and our grid abstraction depicted for P1.\label{fig:dis}}
    \end{subfigure}%
    \hfill
    \begin{subfigure}[t]{0.45\textwidth}
        \centering
        \includegraphics[width=\textwidth]{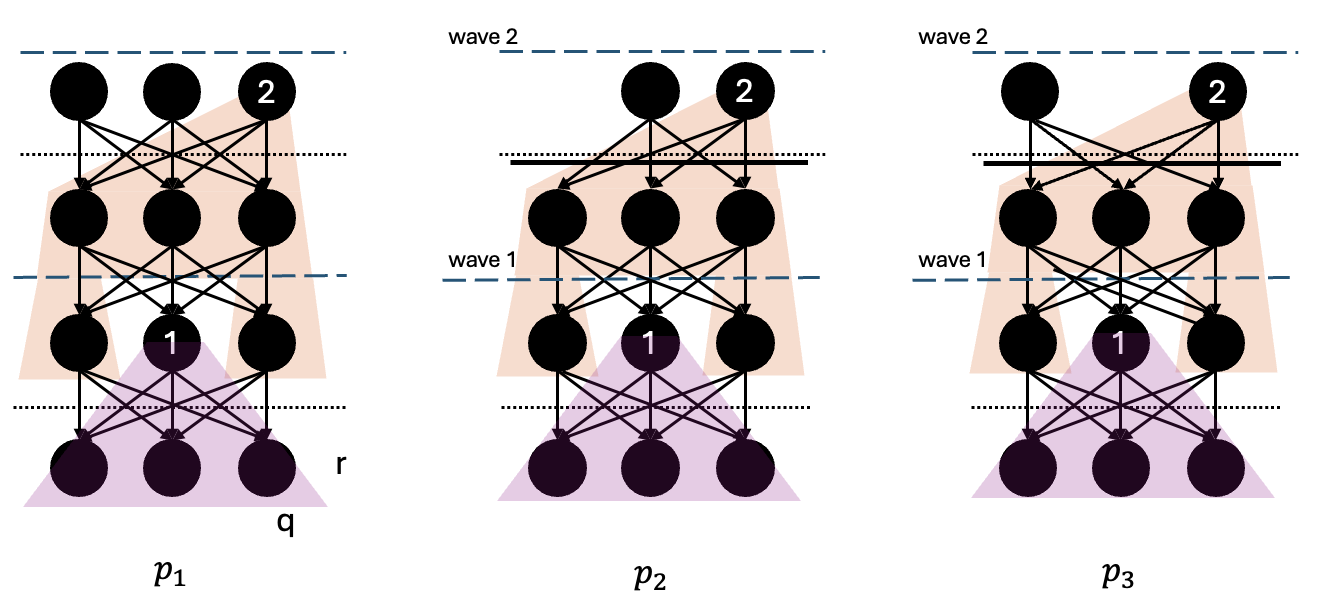}
        \caption{Ordering phase: selecting anchors across local
          DAGs.\label{fig:lin}}
    \end{subfigure}
    \caption{High-level depiction of the two phases in DAG-Rider. The black dots depict vertices that form a local DAG. The dotted blue lines represent a round that ends a wave, and the dotted black line represents a round in a wave.
    }
\end{figure*}

Once a process \( p \) selects anchors for frame \( x \),
it can deliver the vertices from the causal histories of \( A_{(p, x)} \) that have not been delivered before. This procedure builds incrementally on prior frames: before delivering any vertices for frame \( x \), process \( p \) must have selected the anchors for all earlier frames \( 1, 2, \dots, x-1 \) and already delivered their corresponding vertices.  
The delivery process proceeds as follows: when process \( p \) selects the anchor set \( A_{(p,1)} \) for frame 1, it delivers all vertices in its causal history in a deterministic order. For frame 2, process \( p \) delivers only the vertices in the causal history of \( A_{(p,2)} \) that were not already delivered in frame 1. This pattern continues for subsequent frames—at frame \( x \), process \( p \) delivers the vertices in the causal history of \( A_{(p,x)} \) that were not included in any earlier frame.
For example, in Fig.~\ref{fig:lin}, vertex 1 and vertex 2 serve as the singleton sets of anchors for waves 1 and 2, respectively. The shaded regions between vertex 1 and vertex 2 represent differences in causal history, which are ordered deterministically by a topological sort. The elements of this order are delivered sequentially, starting with the causal history of vertex 1, followed by the remaining causal history of vertex 2.

Provided that the Anchor Agreement Property holds (Property \ref{property2}) and the Consistent Causal History Property holds (Property \ref{property1}), the above procedure ensures no duplicates (Integrity) and consistent order (Total Order) across all correct processes. Indeed, Anchor Agreement and Consistent Causal History ensure that all processes deliver the same set of vertices with respect to same frames. Since each frame is sequentially delivered in a deterministic manner, this guarantees Total Order Property. Furthermore, Integrity is preserved because each vertex is delivered exactly once --- vertices from earlier frames are excluded when processing later frames, preventing duplicates.

In the remainder of this subsection we discuss how different DAG based consensus protocols achieve anchor agreement.

\begin{figure}[t]
    \centering
    \includegraphics[width=1\linewidth]{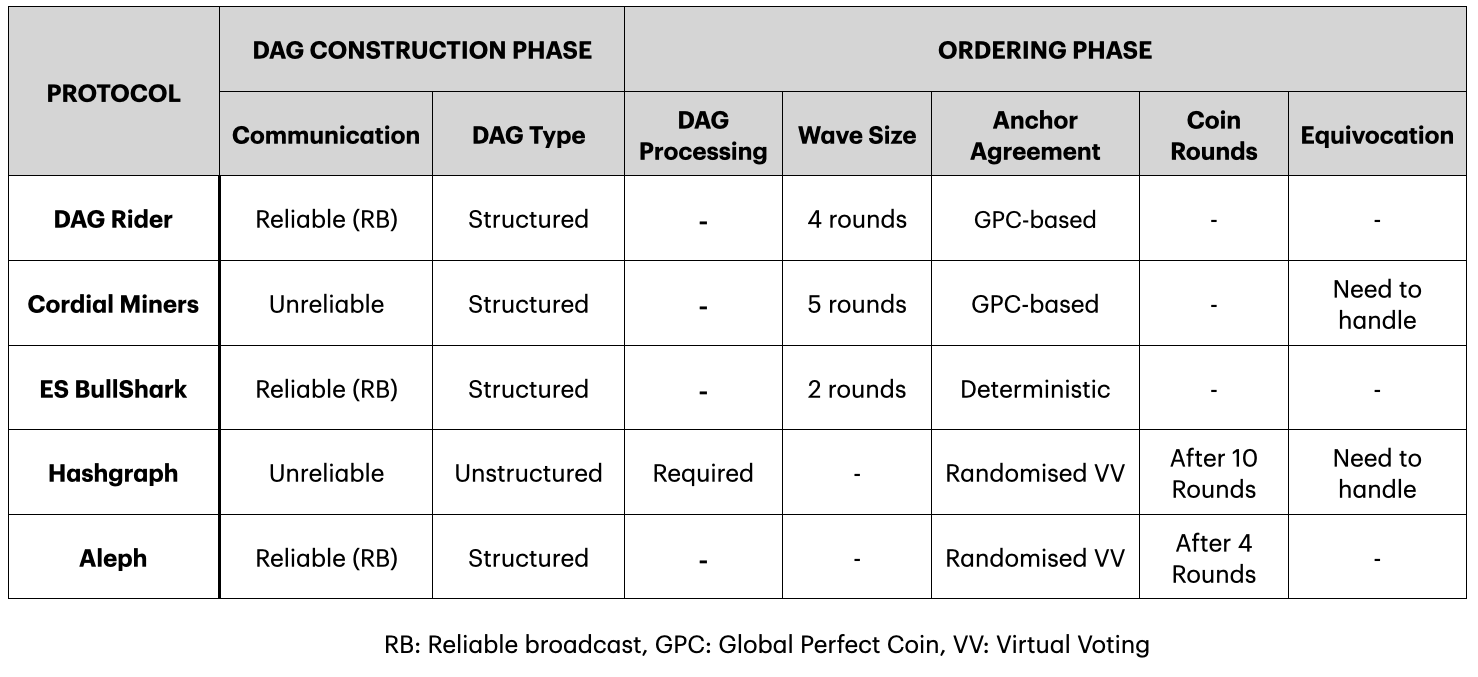}
        \caption{A synthesis of the similarities and variations among different DAG-based consensus protocols.}
    \label{fig:surveyDAG}
\end{figure}

A key property that Anchor Agreement should provide to ensure liveness is that an anchor must eventually appear in the local DAGs of all correct processes. Suppose a correct process \( p \) decides on a vertex \( v \) as one of the anchor vertices for frame \( x \), i.e., \( v \in A_{p,x} \), but \( v \) never appears in the local DAG of a correct process \( q \). Then, correct process \( q \) will never select the anchors of frame \( x \), as \( v \) will never appear in its local DAG. Selecting anchors while excluding \( v \) would violate the Anchor Agreement Property. This prevents the delivery of any vertex from frames bigger than \( x \).
Thus, simple strategies for choosing anchors are insufficient in an asynchronous network. For instance, always selecting the set of vertices created by the process with the lowest ID in the local DAG as anchors might seem plausible. However, this approach fails because an anchor must exist in the local DAGs of all correct processes; having a vertex in one's local DAG does not guarantee its presence in the local DAGs of others. 
To address this, one needs to solve an instance of Binary Agreement to determine whether a vertex exists in the local DAG of every correct process. Due to the Fischer, Lynch, and Paterson (FLP) impossibility result~\cite{FLPjacm85}, which states that no deterministic protocol can guarantee consensus (including binary agreement) in an asynchronous system with faults, randomization must be incorporated into anchor agreement (to ensure progress). In the literature, there are two main variations, discussed below and summarized in Table~\ref{fig:surveyDAG}.

\paragraph{Global Perfect Coin (GPC) Ordering.}\label{background:GPCo} In protocols such as DAG-Rider, Cordial Miners, and BullShark, each process divides its local DAG into a fixed number of consecutive rounds, called \emph{waves} ($4$ in DAG-Rider, $5$ in Cordial Miners, and $2$ in BullShark). Waves act as frames for anchor agreement.
Processes then use a Global Perfect Coin (GPC)~\cite{BandarupalliBBKR24}, a cryptographic primitive to agree on a random leader process in each wave.\footnote{Eventual Synchronous Bullshark, a variant of Bullshark, assumes a certain degree of synchrony in communication. As a result, processes do not have to rely on GPC to compute the leader process; instead, the leader is chosen deterministically, for example, using a round-robin approach.} 
Intuitively, GPC uniformly samples a process ID at random for each wave and incorporates a holding mechanism to enforce unpredictability, especially when only Byzantine processes query it.
The vertices created by the leader process in the first round of each wave are called leader vertices. 
Subsequently, a deterministic leader commit protocol is executed to decide which leader vertex, if any, will be committed as an anchor. 
This commit protocol relies on the properties of the local DAG. 
For instance, protocols that use a reliable broadcast primitive during the DAG construction phase ensure that each process creates a unique vertex in every round, leading to at most one leader vertex per wave in the local DAG.
In contrast, protocols that use unreliable communication cannot guarantee a unique leader vertex in each wave, complicating the agreement on the anchor.

\paragraph{Virtual Voting (VV) Ordering.} \label{background:VVo}
Protocols such as Hashgraph and Aleph simulate a round-based Binary Agreement Protocol (BAP) to determine whether a vertex $v$ from a frame in the local DAG becomes part of the anchor set. Round-based BAPs enable processes to achieve consensus on a binary decision ($0$ or $1$). Each process participates in asynchronous rounds of voting, starting with an initial vote (initial value) in round $1$. In each round, processes broadcast their votes and wait to receive votes from at least $n-f$ processes. The decision rules applied to the received votes dictate the next round’s vote and, the final decision. Certain rounds use randomization in their decision rules (to ensure progress).

In Virtual Voting (VV) ordering, the votes of each process $p$ in each round $r$ (as required by the round-based BAPs) are computed virtually from $G(p)$, based on the relationship between the vertex $v$ (the anchor to be decided) and the vertices created by $p$. Protocols that construct structured DAGs can directly compute these necessary virtual votes. In contrast, protocols that construct an unstructured DAG must first convert them into a structured DAG, referred to as the witness DAG (inspired by \cite{BL-coins20}), 
before computing these votes. The frames used for anchor agreement then correspond to the rounds of the structured DAG. (e.g., the ``local DAG'' in Aleph or the ``witness DAG'' in Hashgraph).

\section{Abstractions Towards Simple and Generic Specifications}
\label{sec:abs}

In this section, we describe the overall architecture of our specifications and the abstractions employed to maximize proof reuse for verifying Properties \ref{property1} and \ref{property2} for the five protocols: DAG-Rider, Cordial Miners, ES Bullshark, Hashgraph, and a variant of Aleph.

\begin{figure}[t]
    \centering
    \begin{minipage}{0.5\textwidth} 
        \centering
        \includegraphics[width=\textwidth]{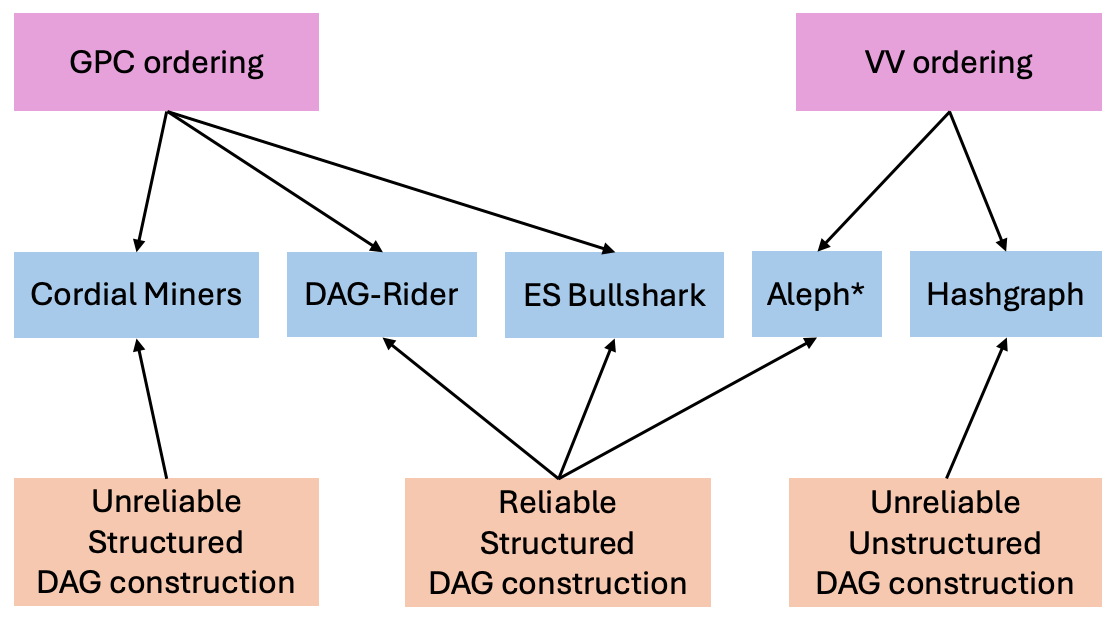} 
    \end{minipage}%
    \hfill
    \begin{minipage}{0.45\textwidth} 
        \caption{Building blocks and their usage in verifying the 5 DAG-based consensus protocols. Each block's/protocol’s TLA+ specification can be found in the corresponding \texttt{Specification.tla} file, and proofs in the corresponding  \texttt{Proofs.tla} file localted in \cite{anonrepo}.}
        \label{fig:buildingblocks}
    \end{minipage}
\end{figure}

First, we divide each protocol specification into two modular components: DAG construction and ordering. This separation reflects the inherent structure of DAG-based consensus protocols (Section~\ref{sec:background}) and facilitates modular verification. A naive such approach to verifying five protocols would require $5 + 5 = 10$ verified specifications. However, our analysis in Section~\ref{background:DAGcon}, summarized in Figure~\ref{fig:surveyDAG}, highlights overlapping patterns in the DAG construction phases. Specifically, the DAG construction phase of DAG-Rider, ES Bullshark, and Aleph is identical. 
By leveraging these similarities, we devise $3$ distinct DAG construction specifications (Figure~\ref{fig:buildingblocks}): (1) Reliable structured, (2) Unreliable structured, and (3) Unreliable unstructured. We discuss the abstractions in them in Subsection~\ref{abs:DAGcon}.

Next, we consolidate the five ordering variations discussed in Section \ref{background:Ordering} and summarized in Figure~\ref{fig:surveyDAG} into two fundamental specifications: GPC ordering and VV ordering. 
This is achieved by isolating the key functionalities of the ordering phase and integrating them into the DAG construction specifications. A detailed discussion is provided in Subsection~\ref{abs:ordering}.

Finally, we refine the ordering specifications by incorporating the DAG construction specifications to produce complete verified specifications for the five protocols (Figure~\ref{fig:buildingblocks}). This refinement step is elaborated in Subsection~\ref{abs:completespec}.
We remark that in Figure~\ref{fig:buildingblocks}, our complete specifications for ES Bullshark and DAG-Rider rely on the same DAG construction and ordering specifications. The differences arise in the refinement step, which defines the wave size, leader position for each wave, and constraints for wave commitment. 

In the remainder of the section,  we write ``GPC ordering specification'' to refer to the TLA+ specification of GPC ordering as found in the file \sloppy \texttt{GPCorderingSpecification.tla}, and similarly for other specifications.

\subsection{Abstractions in DAG Construction Specifications} \label{abs:DAGcon}

\paragraph{Causal Histories as Vertices.}
In our DAG-construction specifications, a vertex in a given position represents not only a block but also its entire causal history within the DAG  (inspired by~\cite{KKNS-podc21}). Listing \ref{lst:datastructures} describes the set of vertices (\texttt{VertexSet}). 
This abstraction significantly simplifies proofs by ensuring that the DAG construction trivially satisfies the Consistent Causal History Property (Property \ref{property1}), which in turn guarantees that the Anchor Agreement Property (Property \ref{property2}) immediately implies agreement on their causal histories, as required to ensure Total Order and Integrity (see Section~\ref{background:Ordering}).

\paragraph{Structured-DAG Data-Structure.}
In our reliable structured and unreliable structured DAG construction specifications, a structured DAG is modeled as a two-dimensional array where the index $(q, r)$ stores vertices $v$ created by process $q$ in round $r$. We refer to index $(q, r)$ as the vertex $v$'s \emph{position}.
Listing \ref{lst:datastructures} describes the type of local DAGs for every process (\texttt{dag}).

\begin{lstlisting}[caption= Some of the data structures in the DAG-construction specifications., label={lst:datastructures}]

VertexSet==[creator:ProcessSet, block:BlockSet, references:SUBSET(VertexSet)]

dag \in [ProcessSet->[RoundSet->[ProcessSet->SUBSET(VertexSet)]]]

broadcastNetwork \in [ProcessSet \cup {"History"}->SUBSET(VertexSet)]
broadcastRecord \in [ProcessSet->[RoundSet -> BOOLEAN]]
\end{lstlisting}

\paragraph{Abstraction of Reliable Broadcast.}
Reliable broadcast ensures three properties (Section~\ref{background:DAGcon}): validity, agreement, and integrity. While validity and agreement are essential for proving consensus liveness, only integrity is required for safety. 
Therefore, we weaken the reliable broadcast abstraction so that it only guarantees integrity, omitting the guarantees of validity and agreement. This weakening 
does not imply that the proofs fail when validity and agreement hold; rather, it ensures the proofs remain applicable even if these properties do not hold. 

Our specification of reliable broadcast also assumes that only the first broadcast vertex of a process in round $r$ can be delivered, thus ensuring integrity.
although this weakens the integrity property, we argue that it is reasonable in this context. Notably, only faulty processes send multiple messages in a single round. If a faulty process broadcasts several messages and one of the later messages is to be delivered, this scenario can be equivalently modeled as an execution where the faulty process sends the later message first. 
 
In our weakened specification, reliable broadcast is modeled using two state variables as shown in Listing \ref{lst:datastructures}. The variable \texttt{broadcast-Network} maintains the sets of vertices to be delivered to each process. Meanwhile, \texttt{broadcast-Record} maintains a log of processes that have broadcast a vertex in each round. A vertex $v$ broadcast in round $r$ is added to \texttt{broadcast-Network} only if \texttt{broadcast-Record} confirms that the same process has not previously broadcast in round $r$. Vertices in the \texttt{broadcast-Network} are then delivered non-deterministically to every process, ensuring that the delivery happens asynchronously.
\paragraph{Abstraction of Unreliable Communication.}
In our unreliable structured and unreliable unstructured DAG construction specifications, we abstract away the specifics of unreliable communication implementations. A process $p$ receiving a vertex from process $q$ is modeled as $p$ non-deterministically choosing a vertex from $q$’s local DAG and copying it. Listing \ref{lst:unrRec} illustrates when the Receive predicate is enabled. This non-deterministic approach captures all possible inter-leavings that a specific asynchronous communication implementation might produce. This abstraction gives a simpler and more generic specification as there are various ways to implement a unreliable communication \cite{gossip}.

\begin{lstlisting}[caption=  Receive in unreliable communication., label={lst:unrRec}]

\* Process p receiving vertex v of round r from process q.

Receive(p, r, v, q) == 
    /\ v \notin dag[p][r][v.creator]
    /\ v \in dag[q][r][v.creator]
    /\ dag' = [ dag EXCEPT ! [p][r][v.creator] = 
            dag[p][r][v.creator] \cup {v}]
\end{lstlisting}

\paragraph{Modeling of Byzantine Behavior.}
A key behavior of Byzantine-faulty processes is their ability to send arbitrary messages. In our setting, since all messages are represented as vertices, we restrict this behavior to allow Byzantine processes to send any vertex. To represent this in reliable communication, we introduce a transition called \texttt{FaultyBcast}, which is specific to faulty processes. 
This transition imposes no constraints on the vertex that can be broadcast, thereby capturing the unpredictable behavior of Byzantine faults. 
In the context of unreliable communication, where the specifics of the underlying communication implementation are abstracted, we model Byzantine behavior by allowing a faulty process to add any vertex to its local DAG, which can then be non deterministically copied by other process.

\subsection{Abstractions in Ordering Specifications} \label{abs:ordering}
{In this subsection, we discuss the ideas and abstractions that enable us to consolidate the five ordering variations into two fundamental specifications: GPC ordering and VV ordering.}

\subsubsection{Global Perfect Coin (GPC) Ordering}

\paragraph{Separation of Concerns for Uniform GPC Ordering.}
DAG construction using reliable broadcast ensures a single vertex at each position in the DAG, while unreliable DAG construction may result in multiple vertices per position. This follows from the Integrity property of reliable broadcast, as discussed in Section~\ref{background:DAGcon}.
To handle both these cases in a uniform way, and to exploit compositionality, we divide the Property \ref{property2} of agreeing on committed wave leader vertices (see  Section \ref{background:GPCo})
into two sub-properties: \textit{Vertex Agreement}: agreeing on a vertex for each position (to address cases with multiple vertices), and \textit{Position Agreement}: agreeing on committed wave leader positions. 
We push \textit{Vertex Agreement} into the DAG construction specifications (specifically, into reliable and unreliable structured DAG construction), and  \textit{Position Agreement} into the GPC ordering specification. 
The safety of \textit{Vertex Agreement} in reliable and unreliable structured DAG construction specifications is formalized as the state invariants \texttt{DAGConsistency} and \texttt{RatificationConsistency}, respectively, as shown in Listing \ref{lst:safety1}. On the other hand, in GPC ordering specification the state invariant \texttt{LeaderConsistency} formalizes the safety of \textit{Position Agreement} as detailed in Listing \ref{lst:safety1}.

\begin{lstlisting}[caption=  Safety Invariants Part 1., label={lst:safety1}]

\* Safety of reliable DAG construction specification. 
\* Note that the dag data structure has been modified to store a single vertex in each position instead of a set of vertices.

DagConsistency == 
   \A p, q \in ProcessorSet, r \in RoundSet, o \in ProcessorSet: 
     ( /\ p \notin faulty /\ q \notin faulty /\ r \= 0 
       /\ dag[p][r][o] \in VertexSet 
       /\ dag[q][r][o] \in VertexSet ) => dag[p][r][o] = dag[q][r][o]
       
---------------------------------------------------------------------
\* Safety of unreliable DAG construction specification.

RatificationConsistency == 
   \A p, q \in ProcessorSet, r \in RoundSet, o \in ProcessorSet: 
     ( /\ p \notin faulty /\ q \notin faulty /\ r \= 0
       /\ dag[p][r][o].ratifiedVertex \in VertexSet 
       /\ dag[q][r][o].ratifiedVertex \in VertexSet ) 
       => dag[p][r][o].ratifiedVertex = dag[q][r][o].ratifiedVertex
       
---------------------------------------------------------------------
\* Safety of GPC ordering specification.

LeaderConsistency == 
   \A p, q \in ProcessorSet: 
     ( /\ p \notin faulty /\ q \notin faulty 
       /\ decidedWave[p] <= decidedWave[q] ) 
       => IsPrefix(leaderSeq[p].current, leaderSeq[q].current)

\end{lstlisting}

\paragraph{Abstraction of Local DAG into Wave-DAG.}
To simplify the GPC ordering specification, we abstract the local DAGs by focusing only on wave leader positions. Since each wave has a unique leader position (see Section \ref{background:GPCo}), we construct a \emph{wave-DAG} consisting of vertices nominated as wave leaders during the ordering phase. Edges in the wave-DAG represent a relevant relation between wave leaders; either the transitive closure relation of strong edges (DAG-Rider)~\cite{KKNS-podc21} or the ratification relation (Cordial Miners)~\cite{KNPS-disc23}.

As shown in Listing~\ref{lst:datastruc2}, the variable \texttt{LeaderRelation} represents a snapshot of the wave-DAG: For every process $p$ and wave ID $w$, {\texttt{LeaderRelation}} stores the following information: (i) the presence of the leader vertex of wave $w$ in the wave-DAG of $p$, and (ii) the set of all waves $w'$ such that the leader vertex of wave $w$ satisfies the corresponding relation with the leader vertex of wave $w'$. 

\begin{lstlisting}[caption= \texttt{leaderRelation} and GPC data structures., label={lst:datastruc2}]

\* state variable leaderRelation capturing the snapshot of wave-DAG

leaderRelation \in [ ProcessSet->[WaveSet->[exists: BOOLEAN, edges: SUBSET(WaveSet)]]]

---------------------------------------------------------------------
\* GPC as a function parameter to the specification

CONSTANT chooseLeader
chooseLeaderTypeAs == chooseLeader \in [WaveSet->ProcessSet]

\end{lstlisting}

\paragraph{Abstraction of Global Perfect Coin.}  
Global Perfect Coin (GPC) (see Section \ref{background:GPCo}) ensures unpredictability, agreement, termination, and fairness when selecting candidate wave-leaders \cite{KKNS-podc21}.
While unpredictability, termination, and fairness are crucial for proving consensus liveness, agreement alone is sufficient for ensuring safety \cite{KKNS-podc21}.
We relax the guarantees of GPC to make its specification compatible with TLA+, which lacks support for expressing randomness.
This relaxed version of GPC does not guarantee unpredictability and need not imply fairness, however it ensures agreement and termination.
Effectively, this relaxation models a scenario where Byzantine processes have increased power, yet the protocol's safety remains uncompromisable.
By removing the requirements of unpredictability and fairness, we represent the global perfect coin as an arbitrary function \texttt{chooseLeader}, shared among all processes, mapping waves to processes, as demonstrated in Listing \ref{lst:datastruc2}. 

\subsubsection*{Virtual Voting (VV) Ordering}

\paragraph{Separation of Concerns for Uniform VV Ordering.}
The reliable structured DAG construction ensures the following (see Section~\ref{background:DAGcon}): (1) a round-based structured local DAG and (2) a property we identify and call \emph{reference consistency}, i.e., if $v$ and $v'$ are round-$r$ vertices in the local DAGs of two correct processes, and if $v$ references $v_p$ and $v'$ references $v_p'$, where $v_p$ and $v_p'$ were created in round $r-1$ by some process $p$, then $v_p = v_p'$. In contrast, unreliable and unstructured DAG constructions neither provide a round-based structured local DAG nor guarantee reference consistency. Since a vertex's vote is determined by its references, one of the requirements for underlying Byzantine Agreement Protocols (BAPs), as in Hashgraph and Aleph, is to construct a structured DAG that ensures reference consistency.
To handle both cases uniformly and leverage compositionality, we divide the task of agreeing on anchors (ensuring Property \ref{property2}) into two sub-tasks: (1) constructing a structured DAG whose vertices satisfy reference consistency, and (2) agreeing on anchors for each frame of the structured DAG, assuming its vertices satisfy reference consistency. 
The specification of (1) is pushed into the DAG construction specifications, while the specification of (2) forms the VV ordering specification.

\begin{lstlisting}[caption= \texttt{WitnessSet} and \texttt{witnessDAG} data structures., label={lst:datastruc4}]

WitnessSet == [source:ProcessSet, frame:FrameSet, vertex:VertexSet, stronglysees:SUBSET(WitnessSet)]

witnessDAG \in [ProcessSet->[Frames->[ProcessSet->SUBSET(WitnessSet)]]]

\end{lstlisting}

In the reliable structured DAG construction specification, the safety invariant \sloppy \texttt{ReferenceConsistency}, defined in Listing \ref{lst:safety2}, formalizes the reference consistency property for the local DAG. 
In the unreliable structured DAG specification, processes constructs a separate structured DAG, whose vertices are called ``witnesses'', referred to as the ``witness-DAG''.
Listing \ref{lst:datastruc4} describes the definitions for the set of witnesses (\texttt{WitnessSet}) and type of the witness-DAG. The safety invariant \texttt{StronglySeenConsistency}, also defined in Listing \ref{lst:safety2}, formalizes the references consistency property for the witness DAG.
Finally, the state invariant \texttt{FameConsistency} formalizes the safety requirements of VV ordering specification, as described in Listing \ref{lst:safety2}.

\begin{lstlisting}[caption= Safety Invariants Part 2., label={lst:safety2}]

\* Safety property of reliable structured DAG construction, it is not hard to see that this follows from earlier defined safety in Listing. 

ReferenceConsistency == 
  \A p \in ProcessSet, q \in ProcessSet, s \in VertexSet, l \in VertexSet: 
     ( /\ p \notin faulty /\ q \notin faulty
       /\ s \in dag[q][s.round][s.creator] 
       /\ l \in dag[p][l.round][l.creator] )
       => ( \A a \in s.referneces, e \in l.referneces: 
            a.round = e.round /\ a.creator = e.creator => a = e )
            
---------------------------------------------------------------------
\* Safety of unreliable unstructured DAG construction. The invariant is rephrased version of TLA spec for clarity, however posses same meaning.

StronglyseenConsistency == 
  \A p \in ProcessSet, q \in ProcessSet, s \in WitnessSet, l \in WitnessSet: 
     ( /\ p \notin faulty /\ q \notin faulty
       /\ s \in witnessDAG[q][s.round][s.creator] 
       /\ l \in witnessDAG[p][l.round][l.creator] )
       => ( \A a \in s.stronglysee, e \in l.stronglysee: 
            a.frame = e.frame /\ a.source = e.source => a = e )
            
---------------------------------------------------------------------
\* Safety of VV ordering specification. DecidedFrame[p][x] is a predicate variable set to true when all leaders of frame x are decided by process p. FamousWitnesses[p][x] stores the set of committed leaders in frame x by process p.

FameConsistency ==
  \A p \in ProcessSet, q \in ProcessSet, x \in Frames:
          DecidedFrames[p][x] /\ DecidedFrames[q][x] 
            => FamousWitnesses[p][x] = FamousWitnesses[q][x]
\end{lstlisting}

To simplify the specifications across all the variants, we abstract away internal states of procedures and focus solely on their input/output behavior. For example, in the $\textit{wave-ready}$ procedure described in algorithm 3 in DAG-Rider \cite{KKNS-podc21}, only the variables $\textit{decidedWave}$, $\textit{deliveredVertices}$, and $\textit{leadersStack}$ are retained, while the loop variable $\textit{w'}$ and the auxiliary variable $\textit{v'}$ are excluded.

\subsection{Complete Protocol Specification} \label{abs:completespec}
For a given protocol, our complete specification refines the ordering specification by incorporating the DAG construction specification. More precisely, we use the TLA+ \emph{interface refinement} construct \cite{Lamport-book2002}, which allows one to obtain a lower-level specification by instantiating the variables of a higher-level specification.
To derive the complete protocol specification, we modify the DAG construction specification in three steps.
\begin{enumerate*}
    \item \emph{Introduce new variables:} For each ordering specification state variable, introduce a corresponding state variable in the DAG construction specification.
    \item \emph{Map variables and constants:} Using the TLA+ \emph{interface refinement} construct, assign values to the ordering specification constants as a function of the DAG construction specification constants. Also, clarify the mapping between the new state variables and the ordering specification state variables.
    \item \emph{Update actions:} For each action in the DAG construction specification, specify how it affects the newly introduced variables using the actions of the ordering specification. If no specific action is performed, define a stutter action where the state variables remain unchanged.
\end{enumerate*}

We illustrate the three steps for DAG-Rider specification, which refines the GPC ordering specification with a reliable structured DAG construction specification.
The GPC ordering specification consists of three constants (\texttt{NumWaves, NumProcesses, Numfaulty}), five variables (\texttt{commitWithRef, decidedWave, leaderRelation, leaderSeq, faulty}), and two actions (\texttt{CreateNewWave, DecideWave}). 
Listing \ref{lst:instanceb} describes the new variables in reliable structured DAG construction specification, their mapping to the variables of the GPC ordering specification, and the assignment to the constants.  When a vertex is added at the leader position of a wave, the \texttt{CreateNewWave} action updates the newly introduced variables. Similarly, when a wave is completed and can be decided in the local DAG, the \texttt{DecideWave} action updates the newly introduced variables. For all other actions in the reliable structured DAG construction, the newly introduced variables remain unchanged.  

\begin{lstlisting}[caption= Instantiation of GPC ordering phase in protocol specification., label={lst:instanceb}]{Name}

VARIABLES commitWithRefN, decidedWaveN, leaderRelationN, leaderSeqN, faultyN

GPCOrdering == INSTANCE GPCOrderingVerification 
               WITH NumWaves <- w,
                    NumProcesses <- n,
                    Numfaulty <- f,
                    commitWithRef <- commitWithRefN,
                    decidedWave <- decidedWaveN,
                    leaderRelation <- leaderRelationN,
                    leaderSeq <- leaderSeqN,
                    faulty <- faultyN
                    
\* w, n, f are constants in reliable structured DAG construction spec
\end{lstlisting}

\section{Evaluation}
\label{sec:eval}
The effort to understand, specify, and prove all the five protocols required approximately 14  person-months, distributed across 5 people. 
We evaluate the performance of our formal proof using TLA+ and TLAPS. 
We direct readers interested in background on TLA+, TLAPS, and proof strategies to Appendix \ref{apd:proofs}. 
Table~\ref{tab:evaluation} lists metrics, including the sizes of the DAG construction specifications and the  ordering specifications.
It also reports on metrics related to proof complexity, such as the number of invariants, the size of proof files, the maximum depth of the proof, the maximum branching in the tree-like proof structure, and the number of base (i.e., leaf) proof obligations.
Finally, the table displays the verification time for proving the safety of each specification.

\begin{table}[t]
\caption{Specification and proof metrics. Performed on a 2.10 GHz CPU with 8 GB of memory, running Windows 11 and TLAPS v1.4.5.}
\label{tab:evaluation}
{
\footnotesize
\begin{center}
\begin{tabular}{l|c|c|c|c|c}
\hline
Metric $\backslash$ Phase & 
\makecell{Reliable \\ structured} & 
\makecell{Unreliable \\ structured} & 
\makecell{Unreliable \\ unstructured} & 
\makecell{GPC \\ Ordering} & 
\makecell{VV \\ Ordering} \\\hline\hline
Size of spec. (\# loc) & 403 & 160 & 230 & 272 & 136 \\
Number of invariants & 6 & 6 & 7 & 10 & 18 \\
Size of proof (\# loc) & 460 & 594 & 554 & 822 & 2120 \\
Max level of proof tree nodes & 10 & 9 & 8 & 9 & 13 \\
Max degree of proof tree nodes & 7 & 8 & 7 & 7 & 11 \\
\# obligations in TLAPS & 633 & 895 & 665 & 1302 & 3316 \\
Time to check by TLAPS (s) & 49 & 68 & 74 & 125 & 651 \\\hline     
\end{tabular}
\end{center}
}
\end{table}
\section{Related Work}
\label{sec:related}
Parameterized verification is the problem of verifying a protocol for arbitrarily many processes~\cite{DBLP:series/synthesis/2015Bloem}. For consensus protocols that tolerate Byzantine failures, this means verifying the protocol is correct for all values of the parameters $n,f$ with $n \geq 3f+1$. 
In comparison, traditional model-checking only handles finite-state systems, and thus finitely many values of the parameters $n,f$. For instance, the Tendermint consensus protocol~\cite{tendermint} has been modeled in TLA+, and certain properties, such as termination, were model-checked for small values of the parameters~\cite{BraithwaiteB0MS20}.
Our work can be seen as a contribution to the parameterized verification of distributed protocols by machine-checkable proofs. An alternative approach is to use parameterized model-checkers, e.g., ByMC~\cite{DBLP:conf/popl/KonnovLVW17}. For instance, 
Safety and Liveness of the consensus algorithm used in the Red Belly Blockchain~\cite{CrainNG21} has been verified for all values of the parameters $n,f$ with $n \geq 3f +1$\cite{redbellyverif}. Although the algorithm only broadcasts binary values, that work verified both safety and liveness (under a weak fairness assumption).

Other BFT consensus protocols have been verified using interactive theorem provers.
The work in~\cite{moonshotverif} uses the IVy interactive theorem prover~\cite{AhnD10} to formally verify a variant of the Moonshine consensus protocol~\cite{moonshotverif}.  
The work in~\cite{LosaD20} uses IVy and Isabelle/HOL~\cite{NipkowPW02} to verify the Stellar Consensus Protocol~\cite{stellar}. 
The Algorand~\cite{algorand} consensus protocol has been verified using Coq~\cite{CoquandH88}.
The safety of several non-Byzantine protocols such as variants of Paxos~\cite{ChandLS16,SchultzDT22} have been verified using interactive theorem proving. 
However, these protocols are not robust to Byzantine faults and they have limited application in blockchains.

To the best of our knowledge, the only other work specific to DAG-based consensus protocols is an unpublished technical report by Crary~\cite{Crary-RR21} describing a formal verification of a DAG-based consensus algorithm, Hashgraph~\cite{BL-coins20}, with the theorem prover Coq.  
In contrast to that approach, our TLA+ specifications of DAG-based consensus protocols are lower-level; they are closer to actual implementations of the protocols and therefore require fewer manual abstractions. Clear advantages of the present approach are its separation of concerns between the communication phase and the ordering phase, and the reusability for several DAG-based consensus protocols.
\section{Conclusion}
\label{sec:conclusion}
Our motivation was to (1) provide a reusable and extensible framework for formally verifying DAG-based consensus protocols, (2) which can subsequently be leveraged to verify their implementations. 
Towards (1), we presented formally verified building-blocks and demonstrated their use in verifying five DAG-based consensus protocols.  To achieve this, we designed the framework with reusability and extensibility in mind, and demonstrated its ability to reduce proof efforts by nearly half across the five protocols we analyzed. 
Other DAG-based consensus protocols can similarly benefit from our framework, even if one just re-uses one of the ordering specifications or DAG construction  specifications. 
Regarding (2), we note that 
machine-checkable proofs can be leveraged to increase trust in implementations. Indeed, generating conformance tests from TLA+ specifications has found previously unknown bugs in distributed systems~\cite{mocket}.

\begin{credits}
\subsubsection{\ackname} The authors gratefully acknowledge support from the Fantom Foundation. This work was also partially supported by the PaVeDyS project (ANR-23-CE48-0005). The authors thank anonymous reviewers for valuable feedback.
\end{credits}

%
%

\bibliographystyle{splncs04}
\bibliography{main}

\end{document}